# Ferroelectric nonlinear anomalous Hall effect in few-layer WTe$_2$

Hua Wang[1] and Xiaofeng Qian[1]*

[1] Department of Materials Science and Engineering, Texas A&M University, College Station, Texas 77843, USA

*Correspondence: Xiaofeng Qian (feng@tamu.edu)

## Abstract

Under broken time reversal symmetry such as in the presence of external magnetic field or internal magnetization, a transverse voltage can be established in materials perpendicular to both longitudinal current and applied magnetic field, known as classical Hall effect. However, this symmetry constraint can be relaxed in the nonlinear regime, thereby enabling nonlinear anomalous Hall current in time-reversal invariant materials – an underexplored realm with exciting new opportunities beyond classical linear Hall effect. Here, using group theory and first-principles theory, we demonstrate a remarkable ferroelectric nonlinear anomalous Hall effect in time-reversal invariant few-layer WTe$_2$ where nonlinear anomalous Hall current switches in odd-layer WTe$_2$ while remaining invariant in even-layer WTe$_2$ upon ferroelectric transition. This even-odd oscillation of ferroelectric nonlinear anomalous Hall effect was found to originate from the absence and presence of Berry curvature dipole reversal and shift dipole reversal due to distinct ferroelectric transformation in even and odd-layer WTe$_2$. Our work not only treats Berry curvature dipole and shift dipole on an equal footing to account for intraband and interband contributions to nonlinear anomalous Hall effect, but also establishes Berry curvature dipole and shift dipole as new order parameters for noncentrosymmetric materials. The present findings, therefore, suggest that ferroelectric metals and Weyl semimetals may offer unprecedented opportunities for the development of nonlinear quantum electronics.

**Introduction**

In classical linear Hall effect, a transverse voltage can be developed in materials with broken time-reversal symmetry only (e.g. in the presence of external magnetic field or internal magnetization) due to Onsager's relation. Second and higher order conductivity tensors, however, are not subject to this constraint, thereby enabling nonlinear anomalous Hall effect (NAHE) in time-reversal invariant system[1-4]. NAHE was observed very recently in few-layer tungsten ditelluride ($WTe_2$)[5-10], a layered material which also holds rich physics including high-temperature quantum spin Hall phase[11-14] and electrostatic gating induced superconductivity[15,16] in its 1T' monolayer and type-II Weyl semimetallicity[17], large non-saturating magnetoresistance[18] and ultrafast symmetry switching[19] in its bulk phase.

Monolayer 1T' $WTe_2$ is centrosymmetric with vanishing even-order nonlinear current response, however vertical electric field can break its two-fold screw rotation symmetry, generate Berry curvature dipole (BCD), and induce second-order nonlinear anomalous Hall current[5-8]. In contrast to monolayer $WTe_2$, bilayer $WTe_2$ is naturally noncentrosymmetric due to the loss of two-fold screw rotation symmetry, resulting in intrinsic nontrivial BCD in bilayer $WTe_2$[9,10]. Surprisingly, ferroelectric switching was recently discovered in semimetallic bilayer and few-layer $WTe_2$[20], very unusual as ferroelectricity and semimetallicity normally do not co-exist in the same material[21]. The subtlety lies in the reduced screening along the out-of-plane direction which gives rise to finite out-of-plane ferroelectric polarization while preserving in-plane semimetallic nature. Conductance hysteresis persisting up to 300K shows its great potential for room temperature device application. These recent studies combined reveal a striking feature of noncentrosymmetric few-layer $WTe_2$ – the coexistence of ferroelectricity and NAHE within a single material, enkindling a few fundamentally and technologically important questions: what's the fundamental correspondence between NAHE and ferroelectricity in ferroelectric metals and Weyl semimetals? Compared to ferroelectric semiconductors[22], what are the unique advantages of ferroelectric metals[20] and ferroelectric Weyl semimetals[23,24]?

Here using first-principles approach and group theoretical analysis we show an intriguing ferroelectric nonlinear anomalous Hall effect (FNAHE) in time-reversal invariant few-layer $WTe_2$ ferroelectric semimetals. In particular, while both bilayer and trilayer $WTe_2$ possess switchable ferroelectric polarization, nonlinear transverse Hall current only switches in trilayer $WTe_2$ upon ferroelectric switching. The microscopic origin of FNAHE in trilayer WTe$_2$ is found to be rooted in the reversal of Berry curvature dipole and shift dipole upon ferroelectric transition, which reveals an exciting yet unexplored realm of ferroelectric metals and Weyl semimetals with potential applications in nonlinear electronics.

**Results and Discussion**

**Ferroelectric transition in bilayer and trilayer WTe$_2$**

Both bilayer and trilayer WTe$_2$ were found to exhibit ferroelectric switching, however their transformation is fundamentally different, which plays a key role in their distinct NAHE. Crystal structures of monolayer, bilayer, and trilayer WTe$_2$ are shown in Fig. 1. Monolayer 1T' WTe$_2$ has a C$_{2h}$ point group with a mirror plane symmetry $\mathcal{M}_y$ perpendicular to y-axis and a two-fold screw rotation symmetry $C_{2y}$. This leads to inversion symmetry $\mathcal{I} = \mathcal{M}_y C_{2y}$ or $C_{2y}\mathcal{M}_y$. Upon van der Waals T$_d$ stacking, multilayer noncentrosymmetric T$_d$ WTe$_2$ possess mirror plane symmetry $\mathcal{M}_y$ only, but no longer hold $C_{2y}$ symmetry as the rotation axes of different layers are not related by any symmetry operation in the point group, consequently multilayer T$_d$ WTe$_2$ lost inversion center with C$_s$ point group.

Ferroelectric transition pathways of bilayer and trilayer T$_d$ WTe$_2$ are shown in Fig. 2. In both cases, two opposite ferroelectric (FE) states (Fig. 2a,c for bilayer, Fig. 2d,f for trilayer) can switch to each other by a small in-plane shift between adjacent layers along $x$ by $2d_x$ ($d_x\sim20$pm), passing through an intermediate paraelectric (PE) state. The intermediate PE state in bilayer WTe$_2$ (Fig. 2b) has a C$_{2v}$ point group with additional $\{\mathcal{M}_z|\frac{1}{2}a\}$ symmetry, thus its out-of-plane electric polarization $P_z$ vanishes. While the ferroelectric transition is achieved by in-plane $2d_x$ shift, of two FE states are related by a glide plane operation $\{\mathcal{M}_z|t_a\}$ consisting of a mirror symmetry operation followed by a translation along $x$ by a fractional translation $t_a$ where $t_a = \frac{1}{2}a + d_x$. For this reason, we denote the two FE states of bilayer WTe$_2$ by -$m$FE and +$m$FE (Fig. 2a,c). In contrast, the two opposite FE states in trilayer WTe$_2$ are related by an inversion operation $\mathcal{I}$, thus denoted by -$i$FE and +$i$FE (Fig. 2d,f). Further, its intermediate PE state has a C$_{2h}$ point group with inversion symmetry, hence the out-of-plane polarization $P_z$ of the PE state in trilayer WTe$_2$ vanishes as well.

We calculate total electronic polarization including both ionic and electronic contributions using first-principles density functional theory (DFT)[25,26]: $P_z = \frac{1}{S}\left(\sum_I Q_I \cdot (z_I - R_0^z) - e\int_V \rho(r)(z - R_0^z)\,d^3r\right)$, where $S$ is the in-plane area of the unit cell, $Q$ is ionic charge, $\rho$ is electronic charge density, and $\boldsymbol{R_0}$ is a reference point which is set to the origin of the unit cell in the present case. The calculated total electronic polarization $P_z$ is $\pm 1.67 \times 10^{-2}$ nm·μC/cm² for $\pm m$FE in bilayer, and $\pm 0.81 \times 10^{-2}$ nm·μC/cm² for $\pm i$FE in trilayer. This is in good agreement with experimentally measured vertical polarization in bilayer WTe₂ of ~ $10^4$ e·cm⁻¹ (*i.e.* $1.60 \times 10^{-2}$ nm·μC/cm²)[20]. Additionally, the intermediate PE state has been recently observed in experiments[19]. In brief, the results from the DFT calculations confirmed the ferroelectricity in both bilayer and trilayer WTe₂, however the symmetry relations between the two FE states are very different in the bilayer and trilayer cases, *i.e.* $-m$FE ↔ PE ↔ $+m$FE and $-i$FE ↔ PE ↔ $+i$FE, which is essential for understanding their distinct NAHE upon ferroelectric switching we will discuss shortly.

**Second-order dc current from NAHE**

Consider an oscillating electric field $\boldsymbol{E}(\boldsymbol{r},t) = \boldsymbol{E}(\omega)e^{i(\boldsymbol{k}\cdot\boldsymbol{r}-\omega t)} + \boldsymbol{E}(-\omega)e^{-i(\boldsymbol{k}\cdot\boldsymbol{r}-\omega t)}$ with $\boldsymbol{E}(\omega) = \boldsymbol{E}^*(-\omega)$ (e.g. under AC electric field or upon coherent light illumination), the second-order nonlinear dc current under minimal coupling approximation is given by $j_a^0 = \sum_{bc} 2\sigma_{abc}(0;\omega,-\omega)E_b(\omega)E_c(-\omega)$. In general $j_a^0$ consists of two parts depending on the polarization of electric field/incident light, including linear photogalvanic effect (LPGE) and circular photogalvanic effect (CPGE)[2,27], *i.e.* $j_a^0 = j_a^L + j_a^C$. Moreover, both LPGE and CPGE have intraband and interband contributions as follows,

$$j_a^L = j_{a,\text{intra}}^L + j_{a,\text{inter}}^L \begin{cases} j_{a,\text{intra}}^L = -2\dfrac{e^3}{\hbar^2}\text{Re}\left(\dfrac{\tau}{1-i\omega\tau}\right)\epsilon_{adc}D_{bd}^{\text{intra}}\,\text{Re}\big(E_b(\omega)E_c(-\omega)\big) \\ j_{a,\text{inter}}^L = -2\dfrac{e^3}{\hbar^2}\tau\, D_{a,bc}^{L,\text{inter}}\,\text{Re}\big(E_b(\omega)E_c(-\omega)\big) \end{cases}$$

$$j_a^C = j_{a,\text{intra}}^C + j_{a,\text{inter}}^C \begin{cases} j_{a,\text{intra}}^C = -\dfrac{e^3}{\hbar^2}\text{Im}\left(\dfrac{\tau}{1-i\omega\tau}\right) D_{ab}^{\text{intra}}\,\text{Im}\big(\boldsymbol{E}(\omega)\times\boldsymbol{E}(-\omega)\big)_b \\ j_{a,\text{inter}}^C = -\dfrac{e^3}{2\hbar^2}\tau D_{ab}^{C,\text{inter}}\,\text{Im}\big(\boldsymbol{E}(\omega)\times\boldsymbol{E}(-\omega)\big)_b \end{cases}$$

Here $\tau$ is the relaxation time and $\epsilon_{adc}$ is the Levi-Civita symbol. $D_{ab}^{\text{intra}}$ is the well-known BCD for intraband nonlinear process[2]. $D_{ab}^{\text{C,inter}}$ is BCD for interband process associated with CPGE[9]. $D_{a,bc}^{\text{L,inter}}$ is shift dipole (SD), originated from the simultaneous displacement of wavepacket upon excitation. More specifically, they are given by

$$\begin{cases} D_{ab}^{\text{intra}}(\mu) = \int_{BZ} f_0(\mu) \partial_a \Omega^b = \int_{BZ} [dk] \sum_n f_n(\mu) \, v_n^a(\mathbf{k}) \, \Omega_n^b(\mathbf{k}) \, \delta(\hbar\omega_n(\mathbf{k}) - \mu) \\ D_{ab}^{\text{C,inter}}(\mu,\omega) = \int_{BZ} [dk] \sum_{mn} f_{nm}(\mu) \, \Delta_{mn}^a(\mathbf{k}) \, \Omega_{nm}^b(\mathbf{k}) \, \text{Re}\left(\frac{\tau}{1 - i(\omega - \omega_{mn})\tau}\right) \\ D_{a,bc}^{\text{L,inter}}(\mu,\omega) = \int_{BZ} [dk] \sum_{mn} f_{nm}(\mu) \, R_{mn}^a(\mathbf{k}) \, \{r_{nm}^b, r_{mn}^c\} \, \text{Re}\left(\frac{1}{1 - i(\omega - \omega_{mn})\tau}\right) \end{cases}$$

Here, $\hbar\omega_n(\mathbf{k})$, $v_n^b(\mathbf{k})$, and $f_n(\mu)$ are band energy, group velocity, and chemical-potential $\mu$ dependent Fermi-Dirac distribution, respectively. $f_{nm}(\mu) \equiv f_n(\mu) - f_m(\mu)$, and $[dk] \equiv d^d k / (2\pi)^d$ for $d$-dimension integral. $\hbar\Delta_{nm}^a \equiv v_n^a - v_m^a$ is the group velocity difference between two bands. $r_{nm}^a$ is interband Berry connection or dipole matrix element. $\Omega_{nm}^c(\mathbf{k})$ is the interband Berry curvature between two bands, defined as $\Omega_{nm}^c(\mathbf{k}) \equiv i\epsilon_{abc} r_{nm}^a r_{mn}^b$. $\Omega_n^z(\mathbf{k})$ is the intraband Berry curvature for band $n$, given by $\Omega_n^c(\mathbf{k}) = \sum_{n \neq m} \Omega_{mn}^c(\mathbf{k})$. In addition, $\{r_{nm}^b, r_{mn}^c\} \equiv \frac{1}{2}(r_{nm}^b r_{mn}^c + r_{mn}^c r_{nm}^b)$. $R_{mn}^a$ is shift vector, given by $R_{mn}^a \equiv -\frac{\partial \phi_{mn}(\mathbf{k})}{\partial k^a} + r_{mm}^a(\mathbf{k}) - r_{nn}^a(\mathbf{k})$, where $\phi_{mn}(\mathbf{k})$ is the phase factor of the interband Berry connection and $r_{nn}^a$ is intraband Berry connection. $\Omega_{nm}^b(\mathbf{k})$, $\Omega_n^b(\mathbf{k})$ and $R_{mn}^a$ are all gauge invariant. For linearly polarized incident light/electric field, $E_b = E_c$, hence we denote $D_{ab}^{\text{L,inter}} \equiv D_{a,bc}^{\text{L,inter}}$.

The intraband and interband BCDs ($D_{ab}^{\text{intra}}$, $D_{ab}^{\text{C,inter}}$) as well as SD ($D_{ab}^{\text{L,inter}}$) have the same units of $L^{3-d}$ for $d$-dimensional system. Thus, BCD and SD have units of length in 2D, but become dimensionless in 3D. At the low frequency limit, $\omega\tau \to 0$, hence $\frac{\tau}{1-i\omega\tau} \to \tau$ and CPGE vanishes. In this case, a dc LPGE current remains which is perpendicular to the applied electric field, thereby inducing *static* NAHE. At high frequency, CPGE becomes nontrivial, referred as to *dynamic* NAHE. The NAHE dc current may switch their direction upon certain ferroelectric transition, giving rise to FNAHE.

**NAHE in bilayer and trilayer WTe$_2$ upon ferroelectric switching**

Now we proceed to discuss NAHE in few-layer WTe$_2$, in particular ferroelectric switching of NAHE (*i.e.* FNAHE) in odd-layer WTe$_2$, and reveal the intriguing connection between BCD/SD and ferroelectric order. We compute their electronic structure by first-principles DFT using hybrid exchange-correlation functional with spin-orbit coupling taken into account. Quasiatomic spinor Wannier functions and tight-binding Hamiltonian were obtained by rotating and optimizing the Bloch functions with a maximal similarity measure with respect to pseudoatomic orbitals[28,29]. Subsequently, first-principles tight-binding approach was applied to compute all the physical quantities such as band structures, BCD, SD, Berry curvature etc. More calculation details can also be found in Methods Section.

Electronic band structure of bilayer WTe$_2$ is presented in Fig. 3a, color-coded by intraband Berry curvature $\Omega_n^z(\bm{k})$. It shows bilayer WTe$_2$ is a small gap insulator, and the intraband Berry curvature is odd with respect to $\Gamma$ due to the presence of time-reversal symmetry. The $\bm{k}$-dependent intraband Berry curvature $\Omega_n^z(\bm{k})$ are shown in Fig. 3c,d at two different chemical potentials of $\mu = \pm 50$ meV. Alternatively, one may use the Kubo formula with the sum-over-states approach for Berry curvature (see Supplementary Fig. S2). Similarly, interband Berry curvature $\Omega_{nm}^z(\bm{k})$ with at frequency $\omega = 120$ meV is displayed in Fig. 3e,f for two sets of occupied and unoccupied bands around the Fermi energy, $\Omega_{\text{VBM}-1,\text{CBM}}^z(\bm{k})$ and $\Omega_{\text{VBM},\text{CBM}-1}^z(\bm{k})$, respectively. VBM refers to valence band maximum, and CBM refers to conduction band minimum. The Berry curvature distribution plots confirm the presence of mirror symmetry $\mathcal{M}_y$ and time-reversal symmetry $\mathcal{T}$. Thus, the integral of the intraband Berry curvature over the full Brillouin zone vanishes, and linear anomalous Hall effect is absent. Furthermore, Figure 3b shows the calculated BCD and SD tensor elements – $D_{yz}^{\text{intra}}$, $D_{yz}^{\text{C,inter}}$, and $D_{xy}^{\text{L,inter}}$ – the key physical quantities governing NAHE. It clearly demonstrates the presence of finite BCD and SD and thus NAHE in bilayer WTe$_2$. The calculated BCD varies between 0-0.4Å depending on the chemical potential, which is in nice agreement with the experimental values of 0.1-0.7Å by Kang et al.[10]. Moreover, upon ferroelectric transition between -*m*FE and +*m*FE state, the Berry curvature, BCD and SD remain unchanged, thus nonlinear anomalous Hall current will not switch direction upon ferroelectric transition in bilayer WTe$_2$. Similarly, the our-of-plane spin polarization remains unflipped, while the in-plane spin polarization is expected to reverse (see Supplementary Fig. S3 and S4).

Trilayer WTe$_2$ is quite different from bilayer WTe$_2$. Figure 4a,b show its electronic band structure of -$i$FE and +$i$FE state, respectively. In contrast to the bilayer case, intraband Berry curvature changes sign upon ferroelectric transition. The similar sign change is also evidenced in the opposite **k**-dependent intraband and interband Berry curvature $\Omega_n^z$ and $\Omega_{\text{VBM}-1,\text{CBM}}^z$ as displayed in Fig. 4e-h. Consequently, the sign of BCD and SD flips upon ferroelectric transition between -$i$FE and +$i$FE, demonstrated in Fig. 4c,d. Therefore, in direct contrast to bilayer WTe$_2$, the nonlinear dc current in trilayer WTe$_2$ will switch its direction upon ferroelectric transition. The calculated BCD ranges from 0 to 0.7Å depending on the chemical potential, also in good agreement with experiment[10]. Moreover, there is a clear plateau in $D_{yz}^{C,\text{inter}}$ marked by purple arrow in Fig. 4c. It is originated from the large joint density of state around 120 meV indicated by purple arrow in Fig. 4a, which remains constant when the chemical potential is located between the energy window. It is also worth to note that, like the bilayer case, the integral of Berry curvature of trilayer WTe$_2$ is also zero due to the presence of time-reversal symmetry, hence the linear anomalous Hall effect is absent. Both in-plane and out-of-plane spin polarizations are reversed (see Supplementary Fig. S5 and S6). Finally, the susceptibility of bilayer and trilayer WTe$_2$ will be reversed in the trilayer case only (e.g. Supplementary Fig. S7 for interband LPGE).

The above electronic structure results demonstrate a striking difference between bilayer and trilayer WTe$_2$, that is, nonlinear anomalous Hall current flips its direction upon ferroelectric switching in trilayer WTe$_2$, but remains unchanged in bilayer WTe$_2$.

**Group theoretical analysis of NAHE in bilayer and FNAHE in trilayer WTe$_2$**

Here we provide a group theoretical analysis of NAHE in addition to the above first-principles calculations. Both bilayer and trilayer WTe$_2$ have C$_s$ point group with a mirror symmetry $M_y$. For circularly polarized incident light propagating along z, $\left(\boldsymbol{E}(\omega) \times \boldsymbol{E}(-\omega)\right)_z$, shares the same $A''$ representation as axial vector $R_z$. Therefore, $\Gamma_{j_y} \otimes \Gamma_{R_{x,z}} = A'' \otimes A'' = A'$, suggesting $\Gamma_{j_y} \otimes \Gamma_{R_{x,z}}$ includes total symmetric irreducible representation, and hence the nonlinear CPGE current can be induced along $y$, i.e. perpendicular to the $xz$ mirror plane. Furthermore, $\Gamma_{j_{x,z}} \otimes \Gamma_{R_{x,z}} = A''$, thus no CPGE current can be induced along $x$. In contrast, for linearly polarized incident light/electric field with in-plane polarization, we have $\Gamma_{j_x} \otimes \Gamma_{E_x} \otimes \Gamma_{E_x} = A' \otimes A' \otimes A' = A'$, and $\Gamma_{j_x} \otimes \Gamma_{E_y} \otimes$ $\Gamma_{E_y} = A' \otimes A'' \otimes A'' = A'$, indicating that the LPGE current can be induced along $x$. However, $\Gamma_{j_y} \otimes \Gamma_{E_x} \otimes \Gamma_{E_x} = \Gamma_{j_y} \otimes \Gamma_{E_y} \otimes \Gamma_{E_y} = A''$, thus no LPGE current can be induced along $y$. This leads to a contrasting CPGE- and LPGE-based nonlinear anomalous Hall current in few-layer WTe$_2$ with C$_s$ point group, that is, linearly polarized light/electric field with in-plane polarization will generate nonlinear anomalous Hall current along $x$ only ($j_x^L \neq 0, j_y^L = 0$), while circularly polarized light propagating along $z$ axis will generate nonlinear anomalous Hall current along $y$ only ($j_x^C = 0, j_y^C \neq 0$).

The correlation between the irreducible representations of parent group C$_{2h}$ and its noncentrosymmetric subgroups C$_2$, C$_s$, and C$_1$ is summarized in Supplementary Table S1. We start from monolayer 1T' WTe$_2$ which has point group of C$_{2h}$, whose second order nonlinear current response vanishes due to the presence of inversion symmetry. Upon vdW stacking (e.g. few-layer and bulk T$_d$ WTe$_2$), $C_{2y}$ is broken with $M_y$ left unchanged, which breaks the inversion symmetry and results in subgroup C$_s$. Consequently, as we analyzed above, $j_x^C = 0$, but $j_y^C \neq 0$ under circularly polarized light, while $j_x^L \neq 0$ but $j_y^L = 0$ under linearly polarized light/electric field with in-plane polarization. However, if $M_y$ is broken with $C_{2y}$ being preserved, it will fall into subgroup C$_2$. In this case, $j_x^C \neq 0$ and $j_y^C = 0$ under circularly polarized light, while $j_x^C = 0$ and $j_y^C \neq 0$ under linearly x/y-polarized light/electric field. Further, if both $M_y$ and $C_{2y}$ are broken, it will end up with subgroup C$_1$, and enable all possible LPGE and CPGE current responses along different directions.

We now discuss the fundamental difference of NAHE between bilayer and trilayer WTe$_2$ upon ferroelectric switching. A general symmetry operator in Seitz notation is given by $g = \{R|\boldsymbol{t}_R\}$, where $R$ is point group symmetry operation, $\boldsymbol{t}_R$ is a translational vector. A time-reversal antisymmetric pseudovector (e.g. Berry curvature and spin polarization) transforms under operator $g$ as follows, $\boldsymbol{m}'(\boldsymbol{k}) = g\,\boldsymbol{m}(\boldsymbol{k}) = P_R P_T R\,\boldsymbol{m}(\boldsymbol{k})$, where $P_R$ and $P_T$ are spatial and temporal parity associated with $g$, respectively. $P_T = \pm 1$ when $R\boldsymbol{k} = \pm\boldsymbol{k} + \boldsymbol{K}$, where $\boldsymbol{K}$ is multiples of reciprocal lattice vector. For bilayer WTe$_2$, as aforementioned, two ferroelectric states can be related by a glide plane operation $\{\mathcal{M}_z|t_a\}$, where $t_a$ refers to a fractional translation along $x$. Thus, $P_R = -1$, $P_T = 1$, and $(m_x, m_y, m_z)^{+m\text{FE}} = (-m_x, -m_y, m_z)^{-m\text{FE}}$. For trilayer WTe$_2$, the two ferroelectric states are related by an inversion operation $\{\mathcal{I}|0\}$, thus $P_R = P_T = -1$, subsequently $(m_x, m_y, m_z)^{+i\text{FE}} = (-m_x, -m_y, -m_z)^{-i\text{FE}}$. The above two conclusions are applicable to any time-reversal antisymmetric pseudovectors such as Berry curvature and spin polarization. For example, for intraband and interband Berry curvature, $\mathcal{M}_z \Omega^z(k_x, k_y) = \Omega^z(k_x, k_y)$ in bilayer WTe$_2$, and $\mathcal{I}\Omega^z(k_x, k_y) = \Omega^z(-k_x, -k_y) \xrightarrow{\text{TRI}} -\Omega^z(k_x, k_y)$ in trilayer WTe$_2$, indicating that the sign of intraband and interband BCD ($D_{ab}^{\text{intra}}$, $D_{ab}^{\text{C,inter}}$) flips only in trilayer WTe$_2$ upon ferroelectric transition. This is in excellent agreement with the first-principles calculations shown in Fig. 3c-f and Fig. 4e-h. In addition, the in-plane spin polarization switches in both cases, and the out-of-plane spin polarization becomes reversed in trilayer WTe$_2$ while remaining unflipped for bilayer WTe$_2$, which also agrees with the calculations (Supplementary Fig. S3-S6). Different from pseudovectors, polar vector $\boldsymbol{p}$ such as electric polarization and shift vector transforms as follows: $\boldsymbol{p}' = R\boldsymbol{p}$. Therefore, both mirror $\mathcal{M}_z$ and inversion $\mathcal{I}$ operation will lead to vertical polarization reversal, i.e. $p_z' = \mathcal{M}_z p_z = -p_z$ and $p_z' = \mathcal{I} p_z = -p_z$, i.e. the out-of-plane electric dipole flips sign in both bilayer and trilayer WTe$_2$ upon ferroelectric transition. In addition, for in-plane shift vector $R_{mn}^a$ with $a \in \{x, y\}$, $(R_{mn}^a)' = \mathcal{M}_z R_{mn}^a = R_{mn}^a$, and $(R_{mn}^a)' = \mathcal{I} R_{mn}^a = -R_{mn}^a$, indicating that the in-plane shift vector $R_{mn}^a$ and thus SD $D_{yz}^{\text{L,inter}}$ will flip only in trilayer WTe$_2$ upon ferroelectric transition. Consequently, the total $j_x^L$ and $j_y^C$ from CPGE and LPGE will switch direction upon ferroelectric transition, provoking FNAHE in time-reversal invariant semimetals. Moreover, it suggests that the BCD and SD can serve as distinct order parameters for noncentrosymmetric semimetals. Figure 5a presents an illustrative summary of the transformation of Berry curvature, spin polarization, and electric polarization under different symmetry operation, while Figure 5b,c show the ferroelectric switching of nonlinear current in the -$i$FE and +$i$FE state of trilayer WTe$_2$. Upon the out-of-plane polarization switching, nonlinear Hall current $j_x^L$ generated via LPGE switches between -$x$ and +$x$ direction under the same external electric field with in-plane linear polarization. Moreover, nonlinear Hall current $j_x^C$ induced by CPGE switches between -$y$ and +$y$ direction under circularly-polarized light with normal incidence. It's worth to emphasize that the intermediate PE state in bilayer and trilayer WTe$_2$ has noncentrosymmetric C$_{2v}$ and centrosymmetric C$_{2h}$ point group, respectively. Thus, despite that the out-of-plane electric polarization vanishes in both cases, nonlinear anomalous Hall current of the PE state vanishes in trilayer, but remains finite in bilayer.

In conclusion, using first-principles calculations and group theoretical analyses we investigated the NAHE in bilayer and trilayer WTe$_2$ and, more importantly, the underlying microscopic origin of FNAHE (*i.e.* ferroelectric switching of NAHE) in trilayer WTe$_2$. Although both bilayer and trilayer WTe$_2$ exhibit ferroelectric transition with similar electric polarization, they behave very differently in NAHE. In the trilayer case, the nonlinear anomalous Hall current flips direction upon ferroelectric switching due to the reversal of BCD and SD under an effective inversion operation of the two ferroelectric states. In contrast, the two ferroelectric states in bilayer WTe$_2$ are related effectively by a glide plane operation which does not flip the BCD/SD, thus its nonlinear anomalous Hall current will not flip upon ferroelectric switching. In addition, NAHE is expected to vanish in the PE state of trilayer WTe$_2$, but remains nontrivial for the PE state of the bilayer case. The above conclusions are applicable to any even and odd layer WTe$_2$ (except monolayer 1T' WTe$_2$ as it is centrosymmetric with vanishing second order NAHE) as long as the two opposite ferroelectric states have the same relationship as the bilayer and trilayer case. The theoretical approaches presented here can also be applied to other materials such as Weyl semimetals[23,24].

More importantly, our results imply that BCD and SD can serve as new order parameters for noncentrosymmetric materials, which opens up the possibility to explore nonlinear multiferroicity based on the coupling of BCD/SD and ferroelectric order. Ferroelectric metals may be advantageous as their vanishing bandgap will not only bring intraband contributions to nonlinear anomalous Hall current that is absent in semiconductors/insulators, but also significantly enhance the interband contributions due to the reduced gap of nonlinear interband processes. For example, the calculated nonlinear anomalous Hall current from interband LPGE in bilayer and trilayer WTe$_2$ is about one order magnitude higher than that in ferroelectric GeSe[22]. Moreover, FNAHE provides a facile approach for direct readout of ferroelectric states, which, combined with vertical ferroelectric writing, may allow for realizing nonlinear multiferroic memory. In addition, the distinct ferroelectric transformation pathway may provide potential routes to realizing non-abelian reciprocal braiding of Weyl nodes[30]. The present findings therefore present an underexplored realm beyond classical linear Hall effect and conventional ferroelectrics with exciting new opportunities for FNAHE-based nonlinear quantum electronics using ferroelectric metals and Weyl semimetals.

## Methods

### First-principles calculations of atomistic and electronic structure

First-principles calculations for structural relaxation, electric polarization, and quasiatomic spinor Wannier functions were performed using density-functional theory[25,26] as implemented in the Vienna Ab initio Simulation Package (VASP)[31] with the projector-augmented wave method[32]. We employed the generalized-gradient approximation of exchange-correlation functional in the Perdew-Burke-Ernzerhof form[33], a plane-wave basis with an energy cutoff of 300 eV, a Monkhorst-Pack k-point sampling of 6×12×1 for the Brillouin zone integration, and optB88-vdW functional[34] to account for dispersion interactions.

### First-principles electronic structure calculations of NAHE

To compute the NHLE-related quantities, we first construct quasiatomic spinor Wannier functions and tight-binding Hamiltonian from Kohn-Sham wavefunctions and eigenvalues under the maximal similarity measure with respect to pseudoatomic orbitals[28,29]. Spin-orbit coupling is taken into account, and hybrid exchange-correlation energy functional HSE06[35] is employed with the range-separation parameter $\lambda = 0.2$ (see Supplementary Information for more details). Total 112 and 168 quasiatomic spinor Wannier functions were obtained for bilayer and trilayer $WTe_2$, respectively. Using the developed tight-binding Hamiltonian we then compute CPGE and LPGE susceptibility tensor with a modified WANNIER90 code[36] using a dense k-point sampling of 600×600×1 for both bilayer and trilayer $WTe_2$. A small imaginary smearing factor $\eta$ of 0.05 eV is applied to fundamental frequency, and Sokhotski-Plemelj theorem is employed for the Dirac delta function integration. In addition, we tested the range separation parameter $\lambda$ in the hybrid HSE functional. Although the values of Berry curvature etc. can change with respect to $\lambda$, the presence (absence) of Berry curvature switching in trilayer (bilayer) remains the same (see Supplementary Fig. S3 for HSE functional with $\lambda = 0.4$). We also checked the convergence of the k-point sampling by increasing it to 1000×1000×1 (Supplementary Fig. S8). Finally, since few-layer $WTe_2$ is either semimetallic or having very small gap, the dielectric screening is large, thus effect of the Coulombic interaction between electrons and holes is negligible.

**Acknowledgements**

This work was supported by the National Science Foundation (NSF) under award number DMR-1753054. Portions of this research were conducted with the advanced computing resources provided by Texas A&M High Performance Research Computing.

**Author contributions**

X.Q. conceived the project. H.W. and X.Q. developed first-principles tight-binding approach for computing and analyzing nonlinear susceptibility tensor. H.W. performed the calculations. Both H.W. and X.Q. analyzed the results and wrote the manuscript.

**Additional information**

**Supplementary information** is available.

**Competing interests:** The authors declare no competing interests.

**Figures**

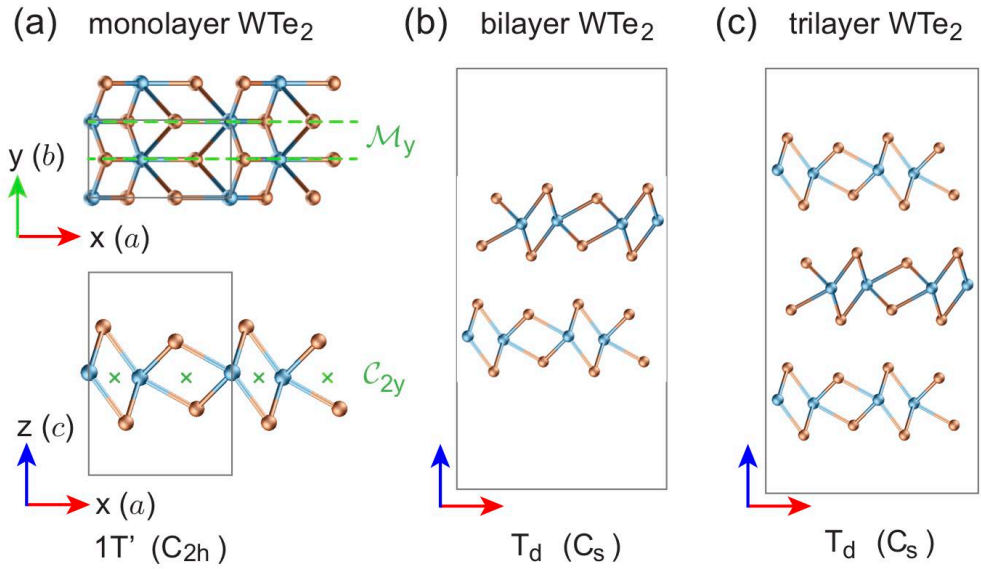

**Fig. 1 Crystal structure of monolayer, bilayer, and trilayer WTe$_2$. a** Monolayer 1T' WTe$_2$ with centrosymmetric C$_{2h}$ point group. It has a mirror plane $\mathcal{M}_y$ and a screw rotation symmetry $C_{2y}$, which leads to the inversion symmetry $\mathcal{I} = \mathcal{M}_y C_{2y}$. **b,c** Bilayer and trilayer T$_d$ WTe$_2$ with C$_s$ point group. $C_{2y}$ symmetry is broken, hence the inversion symmetry $\mathcal{I}$ is also broken in bilayer and trilayer WTe$_2$.

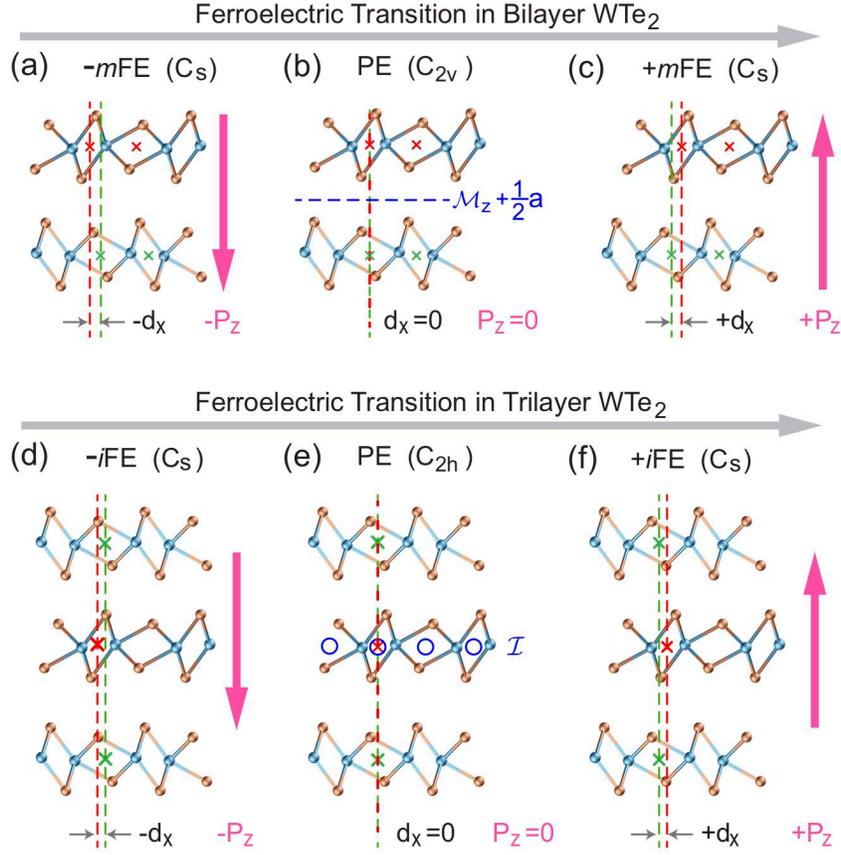

**Fig. 2** Ferroelectric transition in bilayer and trilayer WTe$_2$. **a-c** ferroelectric transition in bilayer WTe$_2$. The two opposite ferroelectric states (-$m$FE and +$m$FE) in bilayer are transformed through a glide plane operation $\{\mathcal{M}_z|t_a\}$, that is, a mirror operation $\mathcal{M}_z$ followed by an in-plane shift along $x$ by $t_a$ with $t_a = \frac{1}{2}a + d_x$. The intermediate PE state C$_{2v}$ point group, thus its out-of-plane polarization with vanishes due to the glide plane $\{\mathcal{M}_z|\frac{1}{2}a\}$ symmetry. **d-f** ferroelectric transition in trilayer WTe$_2$. The two opposite ferroelectric states (-$i$FE and +$i$FE) in trilayer are related to each other through an inversion operation $\{\mathcal{I}|0\}$. The intermediate PE state of trilayer WTe$_2$ has C$_{2h}$ point group, thus its out-of-plane polarization also vanishes due to inversion symmetry. The red and green vertical dashed lines show the relative shift $\pm d_x$ between adjacent WTe$_2$ layers. The corresponding in-plane shift is very small ($d_x \approx 20$pm), therefore it is exaggerated in the above plots for illustrative purpose.

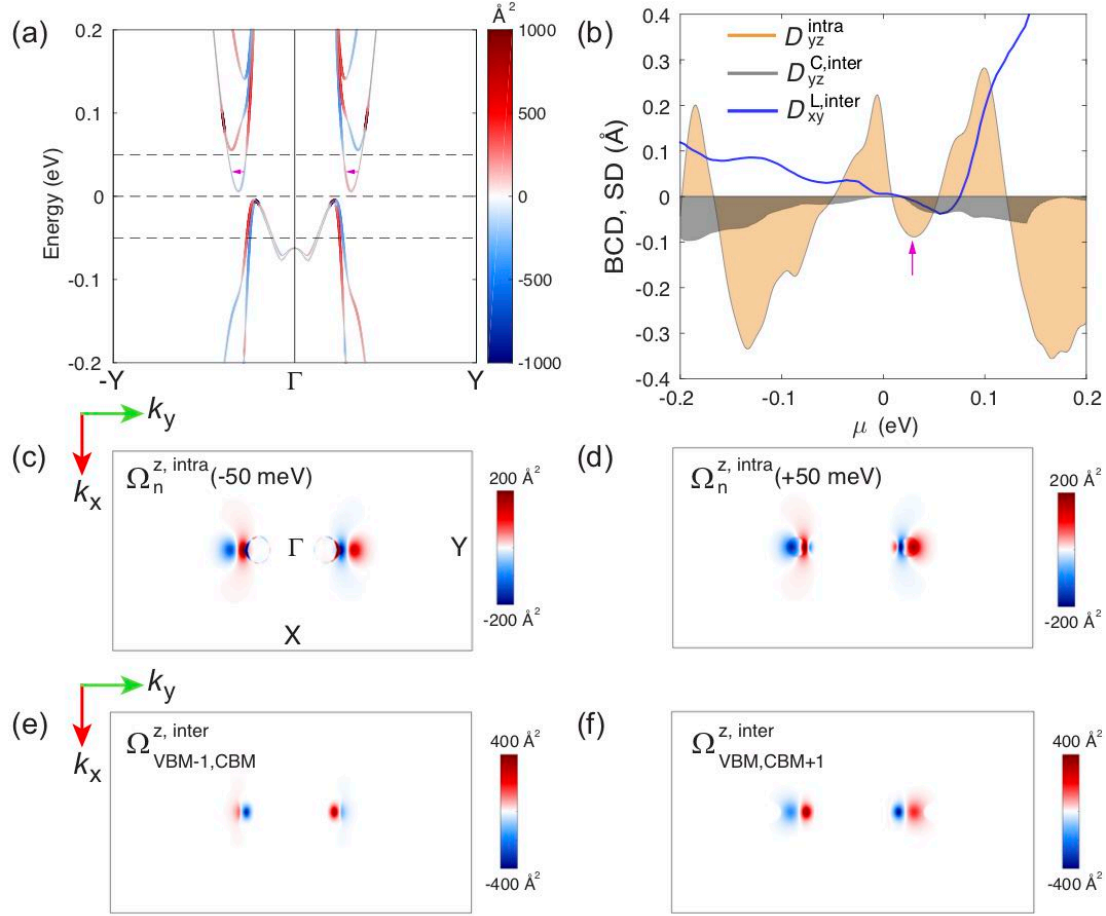

**Fig. 3** Electronic structure and NAHE of bilayer WTe$_2$. **a** Band structure of bilayer WTe$_2$ in $\pm m$FE state color-coded by the z-component of intraband Berry curvature $\Omega_n^z(\boldsymbol{k})$. Spin-orbit coupling is included and hybrid HSE06 functional is employed. **b** BCD and SD tensor elements $D_{yz}^{\text{intra}}(\mu)$, $D_{yz}^{\text{C,inter}}(\mu,\omega)$, and $D_{xy}^{\text{L,inter}}(\mu,\omega)$ as function of chemical potential $\mu$. For interband BCD and SD, $\omega$ is set to 120 meV. **c,d** $\boldsymbol{k}$-dependent distribution of intraband Berry curvature $\Omega_n^z(\boldsymbol{k})$ at $\mu = \pm 50$ meV, respectively. **e,f** $\boldsymbol{k}$-dependent distribution interband Berry curvature $\Omega_{nm}^z(\boldsymbol{k})$ for $(n,m)$=(VBM-1,CBM) and for $(n,m)$=(VBM-1,CBM), respectively. The results are the same for both +$m$FE and -$m$FE state, suggesting that nonlinear anomalous Hall current can be induced in bilayer WTe$_2$, but it will not switch sign upon ferroelectric transition.

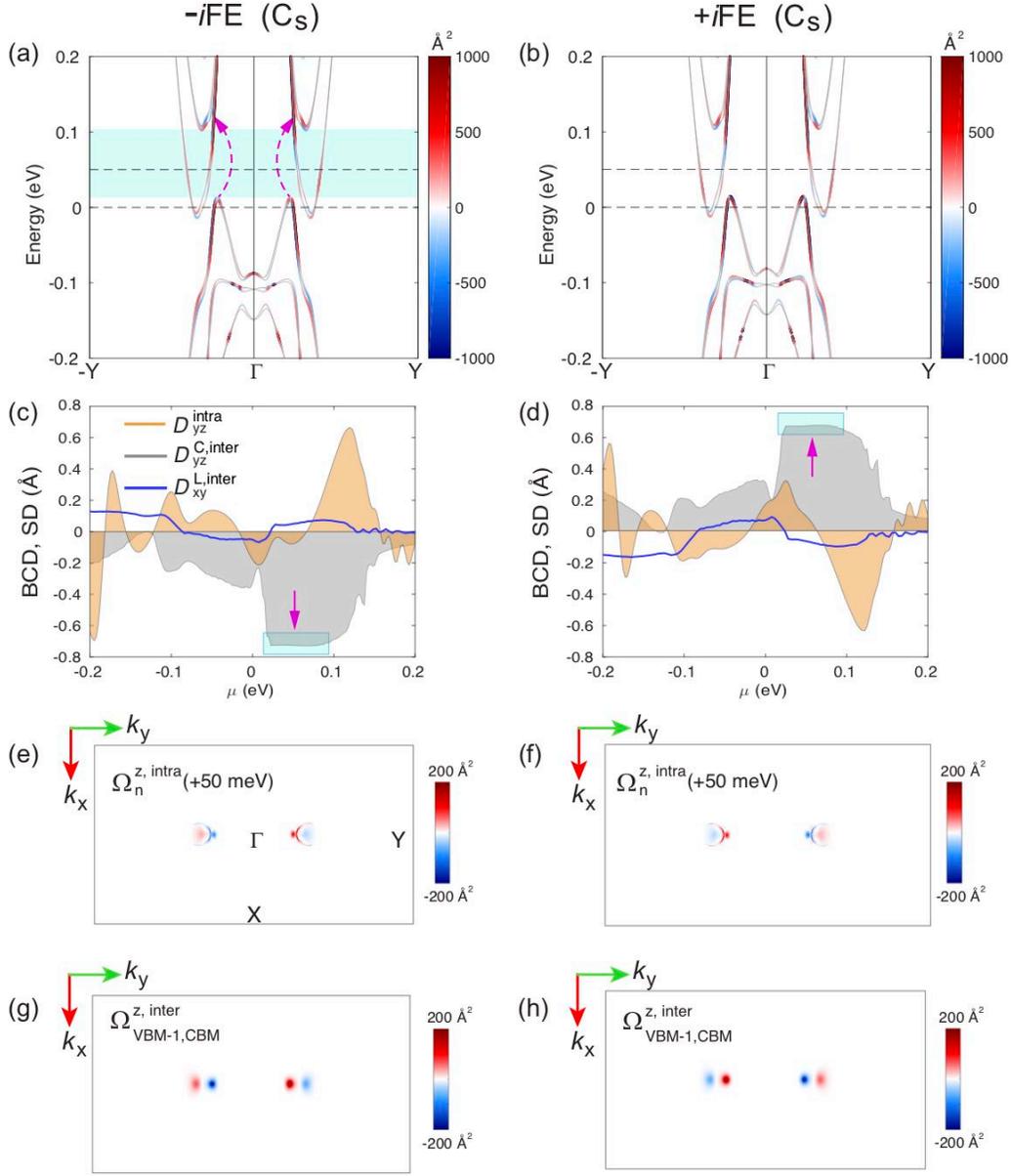

**Fig. 4** Electronic structure and FNAHE of trilayer WTe$_2$. **a,b** Band structure of trilayer WTe$_2$ in -$i$FE and +$i$FE state, respectively. Both are color-coded by the z-component of intraband Berry curvature $\Omega_n^z(\mathbf{k})$. **c,d** BCD tensor elements $D_{yz}^{\text{intra}}(\mu)$ and $D_{yz}^{C,\text{inter}}(\mu,\omega)$, and SD tensor element $D_{xy}^{L,\text{inter}}(\mu,\omega)$ as function of chemical potential $\mu$ for -$i$FE and +$i$FE state, respectively. For interband BCD and SD, $\omega$ is set to 120 meV. **e,f** $k$-dependent distribution of intraband Berry curvature $\Omega_n^z(\mathbf{k})$ at $\mu = \pm 50$ meV for -$i$FE and +$i$FE state, respectively. **g,h** $k$-dependent distribution interband Berry curvature $\Omega_{nm}^z(\mathbf{k})$ between (VBM-1,CBM) around the Fermi surface for -$i$FE and +$i$FE state, respectively. The results clearly show that nonlinear anomalous Hall current in trilayer WTe$_2$ will switch sign upon ferroelectric transition, in direct contrast to the bilayer case.

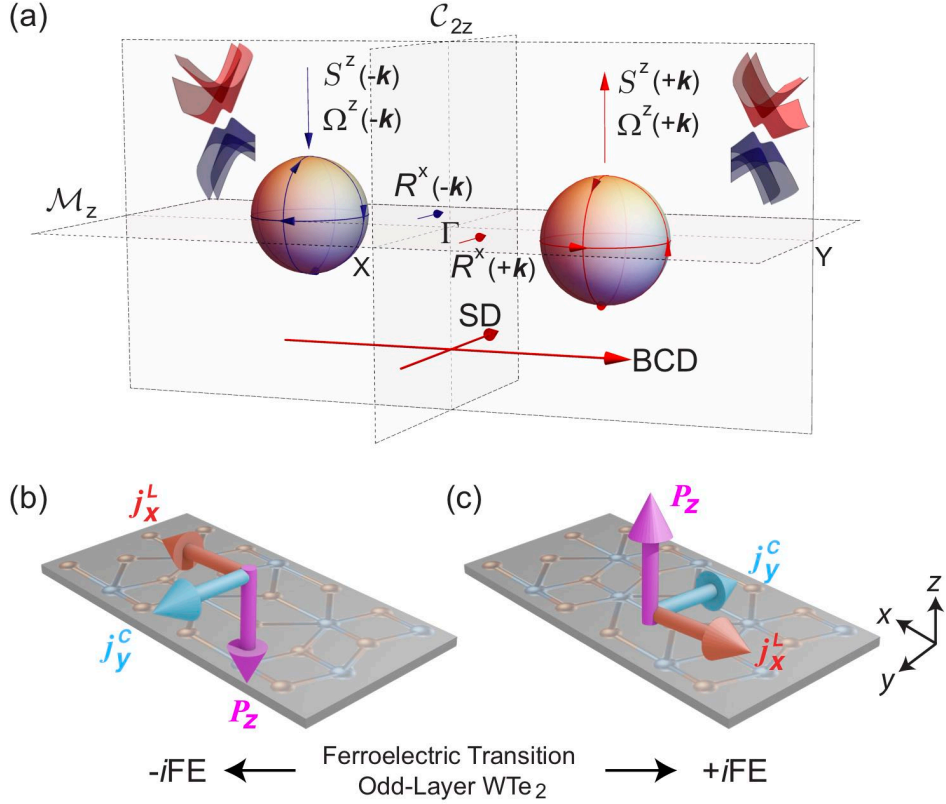

**Fig. 5** Transformation of pseudovectors and polar vectors under different symmetry operation and ferroelectric switching of LPGE and CPGE nonlinear current. **a** Transformation of Berry curvature, spin polarization, shift vector, BCD and SD as well as LPGE and CPGE nonlinear current under different symmetry operation between two ferroelectric states in time-reversal invariant few-layer WTe$_2$. Berry curvature and spin polarization transform as time-reversal antisymmetric pseudovectors. Under a mirror symmetry operation $\mathcal{M}_z$ for the ferroelectric states in bilayer WTe$_2$, most quantities remain invariant, except in-plane spin and Berry curvature component. Under inversion symmetry operation $\mathcal{I}$ for the -$i$FE and +$i$FE state states in trilayer WTe$_2$, all quantities, including Berry curvature, spin polarization, shift vector, BCD and SD, flip the sign in the presence of time-reversal symmetry, giving rise to FNAHE in trilayer WTe$_2$. **b,c** Ferroelectric switching of nonlinear current in the -$i$FE and +$i$FE state of trilayer WTe$_2$, respectively. Upon the polarization $P_z$ switching, nonlinear anomalous Hall current $j_x^L$ from LPGE switches between -$x$ and +$x$ direction under external field with in-plane linear polarization, while nonlinear anomalous Hall current $j_y^C$ from CPGE switches between -$y$ and +$y$ direction under circularly-polarized light with normal incidence.

19